\documentclass[11pt,a4paper]{article}

\usepackage[utf8]{inputenc}
\usepackage[T1]{fontenc}

\usepackage{amsmath}

\usepackage{newtxtext,newtxmath}   
\usepackage{microtype}             

\usepackage{amssymb,amsfonts}

\usepackage[margin=2.5cm]{geometry}
\usepackage{setspace}
\usepackage{graphicx}
\graphicspath{{./}{figures/}{fig/}}
\usepackage{booktabs}
\usepackage{multirow}
\usepackage{caption}
\usepackage{subcaption}
\usepackage{float}
\usepackage{enumitem}
\usepackage{xcolor}

\usepackage{algorithm}
\usepackage[noend]{algpseudocode}

\usepackage{natbib}
\usepackage{url}

\usepackage[colorlinks=true,
            linkcolor=black,
            citecolor=blue,
            urlcolor=blue,
            filecolor=blue]{hyperref}

\usepackage{titlesec}
\titleformat{\section}{\normalfont\large\bfseries}{\thesection}{1em}{}
\titleformat{\subsection}{\normalfont\normalsize\bfseries}{\thesubsection}{1em}{}
\titleformat{\subsubsection}{\normalfont\normalsize\itshape}{\thesubsubsection}{1em}{}

\newcommand{\Addi}{A^{\mathrm{DDI}}}
\newcommand{\Aehr}{A^{\mathrm{EHR}}}
\newcommand{\Aing}{A^{\mathrm{ING}}}
\newcommand{\model}{IngMamba}

\newcommand{\TODO}[1]{\textcolor{red}{\textbf{[TODO: #1]}}}

\newcommand{\safefig}[2][\linewidth]{%
  \IfFileExists{#2}{\includegraphics[width=#1]{#2}}{%
  \IfFileExists{figures/#2}{\includegraphics[width=#1]{#2}}{%
  \IfFileExists{fig/#2}{\includegraphics[width=#1]{#2}}{%
    {\setlength{\fboxsep}{0pt}%
     \fbox{\parbox[c][4.5cm][c]{\dimexpr#1-2\fboxrule\relax}{%
       \centering\ttfamily\small
       [FIGURE PLACEHOLDER]\\[0.6em]
       \detokenize{#2}\\[0.6em]
       file not found}}}%
  }}}%
}

\begin{document}

\title{\textbf{\model: Ingredient-Level Drug--Drug Interaction Modeling\\
with Selective State Space Models for\\
Medication Recommendation on MIMIC-IV}}

\author{
Juao Fan\thanks{Equal contribution.} \and
Jinhan Li\footnotemark[1] \and
Shengxin Zhu\thanks{Corresponding author: \texttt{\TODO{corresponding e-mail}}}
}

\date{%
Guangdong Provincial Key Laboratory of Interdisciplinary Research and Application\\
for Data Science, Beijing Normal--Hong Kong Baptist University, Zhuhai, China\\[1ex]
\today
}

\maketitle

\begin{abstract}
\noindent
Medication recommendation in critical care requires balancing two competing objectives:
clinicians need accurate, personalized prescriptions for complex and evolving patient
needs, while polypharmacy carries a substantial risk of adverse drug--drug interactions
(DDIs). We propose \model, a medication recommendation framework that couples a
selective state space sequence model with multi-granular pharmacological knowledge.
Using the large-scale intensive-care EHR dataset MIMIC-IV, we encode longitudinal
patient trajectories (diagnoses, procedures, and past medications) with a Mamba-based
backbone that handles long, irregular visit sequences in linear time. On top of this
backbone we introduce a joint training objective unifying three knowledge sources:
(i) a drug-level DDI graph, (ii) an ingredient-level DDI graph obtained by normalizing
medication codes to active ingredients via RxNorm, and (iii) an EHR-derived
co-prescription graph reflecting real-world prescribing patterns. This design penalizes
risky combinations while preserving predictive coverage and clinically realistic
co-prescription behavior. Under strictly matched experimental settings---identical
preprocessing, cohort, vocabulary, data split, and evaluation code---\model{}
outperforms a re-implemented MambaHealth baseline on all standard multi-label metrics
(Jaccard $0.4488 \rightarrow 0.4983$, PRAUC $0.6911 \rightarrow 0.7485$,
F1 $0.5989 \rightarrow 0.6453$) while roughly halving the drug-level DDI rate
($0.1875 \rightarrow 0.0948$). We further report an ingredient-level DDI rate, a finer
pharmacological safety measure that is invisible to drug-code-level evaluation. These
results indicate that selective state space modeling and multi-level safety constraints
are complementary rather than competing, and provide a reproducible starting point for
future work on polypharmacy risk reduction in EHR-based clinical decision support.
\end{abstract}

\noindent\textbf{Keywords:} medication recommendation; electronic health records;
MIMIC-IV; state space model; drug--drug interaction; ingredient-level safety;
polypharmacy.

\vspace{1em}

\section{Introduction}

\subsection{Background and Motivation}

Large foundation models---including large language models (LLMs) and multimodal
models---have reshaped general-purpose representation learning, transferring broad
pretraining into downstream tasks with minimal task-specific engineering
\citep{Bommasani2021}. In healthcare, early evidence suggests that LLMs can encode
non-trivial clinical knowledge and support medical reasoning in controlled settings,
although reliability, safety, and evaluation remain central concerns
\citep{Singhal2023}. These trends motivate a concrete question for clinical decision
support: how can we build models that are simultaneously scalable on long,
information-dense clinical trajectories and explicitly constrained by safety knowledge?

Medication recommendation from longitudinal electronic health records (EHRs) is a
representative high-stakes instance of this problem. The task is to recommend a set of
medications for a patient visit given the patient's evolving clinical context. The
difficulty is not only predictive accuracy---matching historically prescribed
regimens---but also safety under polypharmacy: harmful drug--drug interactions (DDIs)
contribute to preventable adverse drug events in hospitalized populations
\citep{deAndradeSantos2020}. Better sequence modeling alone is therefore insufficient;
safety signals must enter both the learning objective and the evaluation protocol.

The data landscape for ICU EHR modeling has matured in parallel. The MIMIC family has
driven open, reproducible critical-care machine learning: MIMIC-III substantially
expanded earlier releases and enabled wide research adoption \citep{Johnson2016MIMICIII},
and MIMIC-IV was subsequently released as a contemporary EHR dataset spanning roughly a
decade of admissions (2008--2019) with more precise digital information sources such as
electronic medication administration records \citep{Johnson2023MIMICIV}.

A practical bottleneck in MIMIC-style medication modeling is vocabulary heterogeneity:
real-world prescription codes (e.g., NDC and RxNorm) are not stable across institutions
and over time. To reduce sparsity and improve interoperability, many clinical pipelines
standardize medication tokens using the WHO Anatomical Therapeutic Chemical (ATC)
classification, which organizes drugs by anatomical, therapeutic, and chemical
properties \citep{WorldHealthOrganizationATC}. This motivates our design choice to unify
medication tokens at a consistent ATC level (ATC3 in our implementation) while retaining
finer pharmacological resolution through ingredient-level safety modeling.

A further observation motivates our central contribution. Drug-code-level DDI graphs
treat each medication token as an indivisible unit, but interactions are mechanistically
determined by \emph{active ingredients}. Two ATC3 tokens may appear unrelated at the code
level while sharing an interacting ingredient pair, and conversely a single code may
aggregate several ingredients with heterogeneous risk profiles. Consequently, safety
constraints imposed purely at the code level are systematically blind to a class of
interaction risks that clinical knowledge bases actually curate at the ingredient level
\citep{nelson2011,Fung2017}.

Overall, this work pursues a single goal: a MIMIC-IV medication recommender that
(i) models long ICU trajectories efficiently and (ii) enforces safety at multiple
pharmacological granularities, rather than treating safety as an afterthought.

\subsection{Contributions}

We propose \model, a safety-aware medication recommendation framework on MIMIC-IV, with
the following contributions.

\begin{itemize}[leftmargin=1.5em,itemsep=0.4em]
\item \textbf{A Mamba-based medication recommendation pipeline for MIMIC-IV with unified
medication tokens.} We construct patient--visit sequences from diagnoses, procedures, and
prescriptions, and standardize medication tokens through an
NDC~$\rightarrow$~RxCUI~$\rightarrow$~ATC4~$\rightarrow$~ATC3 mapping chain to improve
coverage and reduce label sparsity \citep{WorldHealthOrganizationATC}.

\item \textbf{Ingredient-level safety modeling.} Beyond drug-code-level constraints, we
normalize medications to active ingredients via RxNorm \citep{nelson2011} and construct
an ingredient-pair DDI graph from TwoSIDES \citep{tatonetti2012}. Both a training signal
and an evaluation metric are defined at this granularity, matching the level at which
DDI knowledge is curated and applied in practice \citep{Fung2017}.

\item \textbf{A unified multi-granular training objective.} We combine the multi-label
prediction loss with drug-level DDI risk, ingredient-level DDI risk, and an EHR
co-prescription plausibility reward, balanced by a proportional controller that adapts
the safety weight to the observed validation DDI rate.

\item \textbf{Consistent empirical gains under strictly matched settings.} Using a common
pipeline, vocabulary, cohort, split, and evaluation code on MIMIC-IV, \model{} improves
over a re-implemented MambaHealth baseline \citep{Wang2024MambaHealth} on Jaccard, PRAUC,
and F1, while reducing the drug-level DDI rate by approximately $49\%$ relative and
yielding a more clinically plausible average medication count per visit.
\end{itemize}

\section{Related Work}

\subsection{Medication Recommendation on MIMIC-IV}

Medication recommendation predicts a safe and effective medication set for each patient
visit from longitudinal EHRs. With the release of MIMIC-IV as a large-scale, freely
accessible ICU EHR dataset, recent work has re-examined the task under more contemporary
coding, coverage, and cohort settings \citep{Johnson2023MIMICIV}. Earlier longitudinal
models---attention- and memory-based sequence learners---established the importance of
modeling multi-visit trajectories rather than treating visits independently
\citep{Le2018}. Building on this line, more recent methods explore richer temporal
representations for variable-length visit sequences and more structured generation of
medication sets, including hierarchical prediction networks \citep{Liu2023SHAPE} and
conditional generation paradigms for multi-label medication prediction \citep{wu2022}.

\subsection{State Space Models for Long Clinical Sequences}

Transformers have been widely applied to clinical sequence modeling, but their quadratic
cost in sequence length becomes a bottleneck for long EHR histories. Structured state
space models (SSMs) scale more favorably while preserving long-range dependency
modeling, as demonstrated by S4 \citep{Gu2021S4}. Mamba introduces selective state space
modeling to achieve linear-time sequence modeling with strong empirical performance on
information-dense sequences \citep{gu2023}. Outside healthcare, Mamba-style backbones
have been explored for efficient sequential recommendation with competitive
effectiveness--efficiency trade-offs \citep{Liu2024}. Within healthcare, MambaHealth
demonstrated that a lightweight Mamba backbone is viable for drug recommendation on
MIMIC-III \citep{Wang2024MambaHealth}; we adopt it as our baseline and starting point.

\subsection{Safety-Aware Recommendation and DDI Mitigation}

A central challenge in medication recommendation is the trade-off between predictive
accuracy and safety, commonly operationalized through DDI avoidance. A representative
family of methods injects pharmacological knowledge---particularly DDI graphs---into the
learning objective or representation space to penalize high-risk combinations. SafeDrug
introduces molecular-structure-aware representations with explicit DDI-aware learning to
encourage safer combinations while maintaining therapeutic relevance \citep{Yang2021}.
DrugDoctor targets realistic cold-start clinical scenarios by emphasizing visit-level
learning and training strategies that improve early-visit recommendation quality while
retaining safety objectives \citep{kuang2024}.

Closest to our motivation of bridging pharmacology and clinical intent, DATR proposes
DDI-aware therapeutic structure reconstruction, conditionally encoding drug structures
using ATC-derived therapeutic labels and introducing a selectivity-based DDI constraint
to reduce interaction risk during recommendation \citep{Anonymous2025DATR}. The authors
motivate ATC precisely because it links therapeutic intent to chemical structure. This
line of work reflects a broader trend: safe recommendation benefits from aligning
chemical-level signals with clinically meaningful therapeutic organization, rather than
relying on molecular structure alone.

\subsection{Ingredient-Level Pharmacological Modeling}

Despite progress in DDI-aware learning, many safety signals in clinical decision support
are ultimately defined at the active-ingredient level, or require ingredient-normalized
interoperability across drug vocabularies \citep{nelson2011}. Commercial DDI knowledge
bases likewise curate interactions between ingredients rather than between dispensing
codes, and differ substantially in coverage \citep{Fung2017}. Our work advances this
direction by introducing explicit ingredient-level safety modeling in addition to
drug-code-level signals. In contrast to approaches that regularize only code-level
co-prescription, our ingredient-aware design targets interaction mechanisms closer to
pharmacological reality, complementing ATC-based therapeutic organization
\citep{WorldHealthOrganizationATC}.

\subsection{Positioning of \model}

In contrast to prior MIMIC-IV medication recommendation systems that rely on RNN- or
Transformer-style backbones and treat drugs as indivisible tokens, \model{} advances the
literature in two complementary directions. First, it adapts Mamba-style selective SSM
sequence modeling to MIMIC-IV medication recommendation, enabling efficient longitudinal
representation learning over long visit histories. Second, it introduces ingredient-level
modeling as an explicit safety signal, bridging clinical drug codes and pharmacological
mechanism units to complement drug-level DDI constraints. Together, these choices target
both scalability and safety realism for regimen recommendation in complex ICU settings.

\section{Methodology}

\subsection{Data Representation and Medication Token Standardization}
\label{subsec:data}

\paragraph{MIMIC-IV data form.}
For our task, MIMIC-IV is viewed as a collection of relational tables (diagnoses,
procedures, prescriptions) aggregated into patient--visit sequences. Each visit is
identified by the tuple $(\texttt{subject\_id}, \texttt{hadm\_id})$. After aggregation,
each patient becomes a variable-length sequence of visits
\begin{equation}
\mathcal{S}=\{(D_t,P_t,M_t)\}_{t=1}^{T},
\end{equation}
where $D_t$ is the set of diagnosis codes, $P_t$ the set of procedure codes, and $M_t$
the set of medication tokens at visit $t$.

\paragraph{Medication token unification.}
Prescriptions in MIMIC-IV are recorded as NDC codes. To reduce heterogeneity and
sparsity, we standardize medications through the mapping chain
\begin{equation}
\text{NDC} \rightarrow \text{RxCUI} \rightarrow \text{ATC4} \rightarrow \text{ATC3}.
\end{equation}
We use ATC3 as the final medication token space: it provides clinically meaningful
therapeutic grouping while keeping the vocabulary manageable and improving coverage.
Medications that fail to map are dropped rather than backed off to raw NDC/RxCUI, so
that all tokens live in a single consistent space.

\begin{figure}[t]
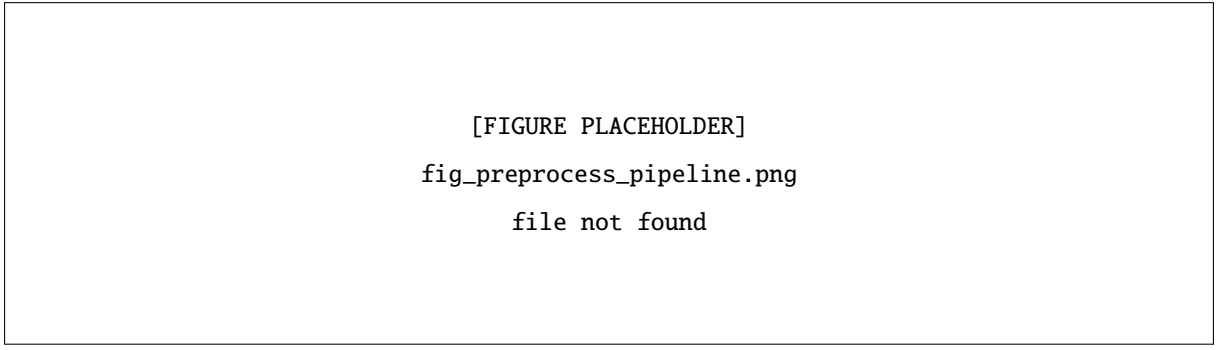

\centering
\safefig{fig_preprocess_pipeline.png}
\caption{MIMIC-IV preprocessing and medication token standardization pipeline
(NDC $\rightarrow$ RxCUI $\rightarrow$ ATC4 $\rightarrow$ ATC3) and generated artifacts.}
\label{fig:preprocess}
\end{figure}

\paragraph{Top-$K$ frequency filtering and outputs.}
To control long-tail noise and stabilize training, we retain the most frequent codes:
the top 2000 diagnoses, top 1000 procedures, and top 300 ATC3 medications. Preprocessing
emits five artifacts: \texttt{data\_final.pkl}, \texttt{voc\_final.pkl},
\texttt{records\_final.pkl}, \texttt{ehr\_adj\_final.pkl}, and \texttt{ddi\_A\_final.pkl}.
Figure~\ref{fig:preprocess} summarizes the pipeline.

\noindent\TODO{add a dataset statistics table here (\#patients, \#visits, avg
visits/patient, avg meds/visit, $|V_d|$, $|V_p|$, $|V_m|$, \#ingredients,
\#ingredient-DDI edges). A skeleton is provided in the commented block below.}


\subsection{Graph Construction for Safety and Clinical Plausibility}
\label{subsec:graphs}

We construct three graphs aligned to the medication vocabulary $V_m$. Each is a symmetric
adjacency matrix $A\in\{0,1\}^{|V_m|\times|V_m|}$ (weighted only during intermediate
counting). Two are described here; the ingredient-level graph $\Aing$ requires additional
normalization machinery and is described in Section~\ref{subsec:ing_graph}.

\subsubsection{EHR Co-prescription Graph ($\Aehr$)}

This graph captures empirical clinical plausibility---what clinicians actually
co-prescribe in the cohort. For each visit $t$, every unordered pair $(i,j)$ with $i\neq
j$ and $i,j\in M_t$ contributes a co-occurrence. Aggregating over all visits,
\begin{equation}
\Aehr_{ij}=
\begin{cases}
1, & \text{if } i \text{ and } j \text{ co-occur in at least one visit},\\
0, & \text{otherwise}.
\end{cases}
\end{equation}
Intuitively, $\Aehr$ provides a data-driven prior over combinations that are compatible
in observed practice.

\subsubsection{Filtered Drug-Level DDI Graph ($\Addi$)}

This graph encodes explicit pharmacological safety constraints. Starting from
\texttt{drug-DDI.csv} (STITCH CID pairs with side-effect annotations) and
\texttt{drug-atc.csv} (CID $\rightarrow$ ATC list), we map
\begin{equation}
(\text{CID}_1,\text{CID}_2) \rightarrow \text{ATC} \rightarrow \text{ATC3}
\rightarrow \text{align to } V_m .
\end{equation}
To improve reliability and reduce spurious density, we apply a three-stage filter:
(i) keep only the top 60 side effects; (ii) keep CID pairs with frequency $\ge 15$;
(iii) cap the result at the 1000 most frequent CID pairs. The resulting $\Addi$ is
substantially sparser and concentrated on higher-confidence risky interactions
(Figure~\ref{fig:graphs}).

\begin{figure}[t]
\centering
\safefig{fig_two_graphs.png}
\caption{Two graphs aligned to the medication vocabulary: (A) the EHR co-prescription
graph $\Aehr$ encoding empirical plausibility, and (B) the filtered DDI graph $\Addi$
encoding explicit safety constraints, retaining the top 60 side effects, pairs with
frequency $\ge 15$, and at most 1000 CID pairs.}
\label{fig:graphs}
\end{figure}

\subsection{Ingredient-Level DDI Graph}
\label{subsec:ing_graph}

\subsubsection{Drug Normalization}

We extract NDC codes from \texttt{prescriptions.csv} in MIMIC-IV, normalize the raw
strings (removing non-digit characters and generating zero-padded candidates), and query
RxNorm \citep{nelson2011} to map each code to its corresponding RxCUI.

\subsubsection{Ingredient Vocabulary}

We follow the \texttt{IN}/\texttt{PIN} (ingredient / precise ingredient) relationships in
RxNorm to extract drug--ingredient mappings, yielding a function
$I(\cdot)$ from \texttt{drug\_rxcui} to a set of \texttt{ingredient\_ids}. All ingredients
are assigned indices in a trainable embedding table.

\subsubsection{Graph Construction and Projection to $V_m$}

Using the TwoSIDES resource \citep{tatonetti2012}, we standardize drug identifiers into
ingredient sets and build an ingredient-pair DDI graph $G$, aggregating support counts,
adverse-drug-event overlaps, and proportional reporting ratio (PRR) statistics into a
sparse adjacency representation.

Two views of this graph are used, and it is important to distinguish them:
\begin{itemize}[leftmargin=1.5em,itemsep=0.2em]
\item \textbf{For training}, we project $G$ onto the medication vocabulary to obtain
$\Aing \in \{0,1\}^{|V_m|\times|V_m|}$, where $\Aing_{ij}=1$ if any ingredient of token
$i$ interacts with any ingredient of token $j$ in $G$. This projection makes $\Aing$
dimensionally compatible with the probability-based safety scores of
Section~\ref{subsec:safety}.
\item \textbf{For evaluation}, we operate in the native ingredient space: a predicted
medication set is decomposed into its ingredient set and pairs are checked directly
against $G$ (Algorithm~\ref{alg:ing_ddi}). This avoids the information loss introduced
by projection and yields the ingredient-level DDI rate reported in
Section~\ref{subsec:safety_metrics}.
\end{itemize}

\subsection{Model Architecture}

\model{} consists of two core modules: patient health state modeling (a two-stream shared
Mamba encoder) and prescription generation (fusion, attention head, and multi-label
output), with safety scores computed from the predicted probabilities
(Figure~\ref{fig:architecture}).

\subsubsection{Visit-Level Set Embedding}

Let $E^d\in\mathbb{R}^{|V_d|\times d}$ and $E^p\in\mathbb{R}^{|V_p|\times d}$ be learnable
embedding tables for diagnoses and procedures. For visit $t$ we sum the embeddings of the
codes in each set,
\begin{equation}
e_t^{d}=\sum_{c\in D_t} E^d[c],\qquad
e_t^{p}=\sum_{c\in P_t} E^p[c],
\end{equation}
yielding two sequences $\{e_t^{d}\}_{t=1}^{T}$ and $\{e_t^{p}\}_{t=1}^{T}$. We use
$d=128$ and dropout $0.3$.

\subsubsection{Two-Stream Shared Mamba Encoder}

We apply $L=3$ stacked Mamba blocks to each stream across time:
\begin{equation}
H^{d}=\mathrm{Mamba}^{(L)}(\{e_t^{d}\}_{t=1}^{T}),\qquad
H^{p}=\mathrm{Mamba}^{(L)}(\{e_t^{p}\}_{t=1}^{T}),
\end{equation}
where $H^{d},H^{p}\in\mathbb{R}^{T\times d}$ are per-visit hidden states. The two streams
share weights, which halves the parameter count of the encoder and empirically stabilizes
training on patients with few visits. We use lightweight Mamba settings
$d_{\mathrm{conv}}=4$ and $\mathrm{expand}=2$, and set the SSM state dimension to
\begin{equation}
d_{\mathrm{state}}=|V_m| .
\end{equation}
This choice was made empirically; we did not observe instability from the resulting state
width, though a systematic sensitivity analysis over $d_{\mathrm{state}}$ is left to
future work.

\subsubsection{Fusion and Attention-Based Prescription Head}

At each visit $t$ we fuse the two streams by concatenation and project back to $d$:
\begin{equation}
h_t=\mathrm{MLP}\!\left([h_t^{d};h_t^{p}]\right)\in\mathbb{R}^{d}.
\end{equation}
We then apply single-head self-attention over the temporal states $H=\{h_t\}_{t=1}^{T}$,
\begin{equation}
Q=HW_Q,\quad K=HW_K,\quad V=HW_V,\qquad
\mathrm{Att}(H)=\mathrm{softmax}\!\left(\frac{QK^\top}{\sqrt{d}}\right)V,
\end{equation}
with a causal mask so that visit $t$ attends only to visits $\le t$. With a residual
connection and layer normalization, the attended representation $\tilde{h}_t$ produces
medication logits
\begin{equation}
z_t=W_o\tilde{h}_t+b,\qquad p_t=\sigma(z_t),
\end{equation}
where $p_t\in(0,1)^{|V_m|}$ is a multi-label probability vector. The predicted medication
set is
\begin{equation}
\hat{M}_t=\{i\mid p_{t,i}\ge \tau\},\qquad \tau=0.5 .
\end{equation}

\begin{figure}[t]
\centering
\safefig{fig_model_architecture.png}
\caption{End-to-end framework and model architecture of \model: a two-stream shared Mamba
encoder for patient state modeling, a fusion and attention-based prescription head, and
probability-based multi-granular safety scoring.}
\label{fig:architecture}
\end{figure}

\subsection{Safety Scoring and Training Objective}
\label{subsec:safety}

\subsubsection{Probability-Based Quadratic Safety Scores}

All three graphs of Sections~\ref{subsec:graphs} and \ref{subsec:ing_graph} enter the
objective through the same differentiable quadratic form applied to the predicted
probabilities $p_t$. For the drug-level DDI adjacency,
\begin{equation}
r_t^{\mathrm{DDI}}
= s\cdot \sum_{i=1}^{|V_m|}\sum_{j=1}^{|V_m|} p_{t,i}\,p_{t,j}\,\Addi_{ij}
= s\cdot p_t^\top \Addi p_t ,
\end{equation}
where $s=5\times10^{-4}$ is a scaling factor keeping the safety terms on a comparable
scale to the task loss. Analogously, the EHR co-prescription plausibility score and the
ingredient-level risk score are
\begin{equation}
r_t^{\mathrm{EHR}}=s\cdot p_t^\top \Aehr p_t ,
\qquad
r_t^{\mathrm{ING}}=s\cdot p_t^\top \Aing p_t .
\end{equation}
Note the difference in sign convention: $r_t^{\mathrm{DDI}}$ and $r_t^{\mathrm{ING}}$ are
\emph{risks} to be minimized, whereas $r_t^{\mathrm{EHR}}$ is a \emph{reward} measuring
agreement with observed co-prescription practice and is therefore maximized.

\subsubsection{Multi-Label Task Loss}

The task loss combines binary cross-entropy with a multi-label margin term
$\mathcal{L}_{\mathrm{MLM}}$, following standard practice in this literature
\citep{Yang2021}:
\begin{equation}
\mathcal{L}_{\mathrm{task}}=\alpha\,\mathcal{L}_{\mathrm{BCE}}
+(1-\alpha)\,\mathcal{L}_{\mathrm{MLM}},\qquad \alpha=0.95,
\end{equation}
where, for a visit with ground-truth set $Y_t$,
\begin{equation}
\mathcal{L}_{\mathrm{BCE}}
= -\frac{1}{|V_m|}\sum_{i=1}^{|V_m|}
\Big[ y_{t,i}\log p_{t,i} + (1-y_{t,i})\log (1-p_{t,i}) \Big],
\end{equation}
\begin{equation}
\mathcal{L}_{\mathrm{MLM}}
= \frac{1}{|V_m|}\sum_{i \in Y_t}\ \sum_{j \notin Y_t}
\max\big(0,\, 1 - (p_{t,i} - p_{t,j})\big).
\end{equation}

\subsubsection{Unified Multi-Granular Objective}

Averaging each safety score over visits gives
\begin{equation}
\mathcal{L}_{\mathrm{DDI}}=\frac{1}{T}\sum_{t=1}^{T} r_t^{\mathrm{DDI}},\qquad
\mathcal{L}_{\mathrm{ING}}=\frac{1}{T}\sum_{t=1}^{T} r_t^{\mathrm{ING}},\qquad
\mathcal{L}_{\mathrm{EHR}}=\frac{1}{T}\sum_{t=1}^{T} r_t^{\mathrm{EHR}} .
\end{equation}
The full objective couples the task loss with the three knowledge sources:
\begin{equation}
\label{eq:total_loss}
\mathcal{L}=\beta\,\mathcal{L}_{\mathrm{task}}
+(1-\beta)\Big(
\lambda_{\mathrm{DDI}}\,\mathcal{L}_{\mathrm{DDI}}
+\lambda_{\mathrm{ING}}\,\mathcal{L}_{\mathrm{ING}}
-\lambda_{\mathrm{EHR}}\,\mathcal{L}_{\mathrm{EHR}}
\Big),
\end{equation}
with $\lambda_{\mathrm{DDI}}=\TODO{value}$, $\lambda_{\mathrm{ING}}=\TODO{value}$, and
$\lambda_{\mathrm{EHR}}=\TODO{value}$.
Setting $\lambda_{\mathrm{ING}}=\lambda_{\mathrm{EHR}}=0$ recovers the drug-level-only
objective used by the MambaHealth baseline, which makes the comparison in
Section~\ref{subsec:results} a controlled one.

\subsubsection{Dynamic Safety Weighting}

Rather than fixing the accuracy--safety trade-off a priori, we adapt $\beta$ with a
proportional controller driven by the validation drug-level DDI rate
$R_{\mathrm{DDI}}$ against a target $\gamma$:
\begin{equation}
\beta_{\mathrm{new}}=\mathrm{clip}\Big(\beta + K_p(\gamma - R_{\mathrm{DDI}}),\;
\beta_{\min},\beta_{\max}\Big),
\end{equation}
with $\gamma=0.05$, $K_p=0.5$, $\beta_{\min}=0.2$, and $\beta_{\max}=0.95$. When the
observed DDI rate exceeds the target, $\beta$ decreases and the safety terms gain weight;
when it falls below, the model is allowed to prioritize accuracy. For stability we apply
exponential moving average smoothing,
\begin{equation}
\beta \leftarrow m\beta + (1-m)\beta_{\mathrm{new}},\qquad m=0.9 .
\end{equation}

\subsection{Training Protocol and Implementation Details}

We split patients (not visits) into train/validation/test sets with an 80/10/10 ratio, so
that no patient trajectory spans two splits. Training uses Adam with learning rate
$3\times10^{-5}$ for up to 50 epochs with early stopping (patience 10) on validation
Jaccard. Mini-batches are formed at the patient-sequence level, and variable-length visit
sequences are preserved by a custom collate function. At inference we threshold at
$\tau=0.5$.

\subsection{Evaluation Procedures}

Algorithms~\ref{alg:drug_ddi} and \ref{alg:ing_ddi} give the two DDI evaluation
procedures. The drug-level procedure is identical to the reference implementation used by
MambaHealth and prior work, ensuring that baseline and \model{} are scored by exactly the
same code path.

\begin{algorithm}[t]
\caption{Drug-level DDI rate for a single visit}
\label{alg:drug_ddi}
\begin{algorithmic}[1]
\Require Predicted medication set $\hat{Y}_v$; drug-level DDI adjacency $A$
\Ensure Drug-level DDI rate for visit $v$
\State $\textit{ddi} \gets 0$; $\textit{pairs} \gets 0$
\ForAll{pairs $(m_i,m_j)$ in $\hat{Y}_v$ with $i<j$}
    \State $\textit{pairs} \gets \textit{pairs} + 1$
    \If{$A[m_i,m_j] = 1$}
        \State $\textit{ddi} \gets \textit{ddi} + 1$
    \EndIf
\EndFor
\Return $\textit{pairs} > 0\ ?\ \textit{ddi}/\textit{pairs}\ :\ 0$
\end{algorithmic}
\end{algorithm}

\begin{algorithm}[t]
\caption{Ingredient-level DDI rate for a single visit}
\label{alg:ing_ddi}
\begin{algorithmic}[1]
\Require Predicted medication set $\hat{Y}_v$; ingredient mapping $I$; ingredient DDI graph $G$
\Ensure Ingredient-level DDI rate for visit $v$
\State $\mathcal{I}_v \gets \bigcup_{m \in \hat{Y}_v} I(m)$
\State $\textit{ddi} \gets 0$; $\textit{pairs} \gets 0$
\ForAll{pairs $(g_i,g_j)$ in $\mathcal{I}_v$ with $i<j$}
    \State $\textit{pairs} \gets \textit{pairs} + 1$
    \If{$(g_i,g_j) \in G$}
        \State $\textit{ddi} \gets \textit{ddi} + 1$
    \EndIf
\EndFor
\Return $\textit{pairs} > 0\ ?\ \textit{ddi}/\textit{pairs}\ :\ 0$
\end{algorithmic}
\end{algorithm}

\section{Experiments}

We validate \model{} from both predictive-accuracy and pharmacological-safety
perspectives. All evaluations use MIMIC-IV under a unified preprocessing pipeline. We
compare (i) our re-implemented MambaHealth baseline and (ii) \model. We additionally
reproduce published numbers from DATR \citep{Anonymous2025DATR} for context; because DATR
uses its own preprocessing and split protocol, those values are reference points rather
than strictly comparable baselines.

\subsection{Evaluation Metrics}

\subsubsection{Accuracy}

For each visit $v$ with ground-truth medications $Y_v$ and predicted set $\hat{Y}_v$,
\begin{equation}
\mathrm{Jaccard}(v) = \frac{|Y_v \cap \hat{Y}_v|}{|Y_v \cup \hat{Y}_v|},\qquad
\mathrm{Precision} = \frac{|Y_v \cap \hat{Y}_v|}{|\hat{Y}_v|},\qquad
\mathrm{Recall} = \frac{|Y_v \cap \hat{Y}_v|}{|Y_v|},
\end{equation}
\begin{equation}
\mathrm{F1} = \frac{2 \cdot \mathrm{Precision} \cdot \mathrm{Recall}}
{\mathrm{Precision} + \mathrm{Recall}} .
\end{equation}
We additionally report PRAUC, the area under the precision--recall curve computed over
the medication probability vector, which is threshold-independent and better suited to
the strong label imbalance of the task.

\subsubsection{Drug-Level DDI Rate}

Following MambaHealth and prior work, the drug-level DDI rate uses the released DDI
matrix \texttt{ddi\_A\_final.pkl}. For visit $v$,
\begin{equation}
\mathrm{DDI}_{\mathrm{drug}}(v)
= \frac{\sum_{i<j} A_{m_i, m_j}}{\binom{|\hat{Y}_v|}{2}},
\qquad
\text{Drug-DDI} = \frac{1}{|\mathcal{V}|}\sum_{v \in \mathcal{V}}
\mathrm{DDI}_{\mathrm{drug}}(v),
\end{equation}
where $\mathcal{V}$ is the set of test visits. This is identical to the computation in the
original MambaHealth implementation, ensuring strict fairness between baseline and \model.

\subsubsection{Ingredient-Level DDI Rate}
\label{subsec:safety_metrics}

To evaluate safety at finer pharmacological resolution, we decompose each predicted
medication set into its active ingredient set $\mathcal{I}_v$ and score pairs against the
ingredient-level DDI graph $G$ (Algorithm~\ref{alg:ing_ddi}):
\begin{equation}
\mathrm{DDI}_{\mathrm{ing}}(v)
= \frac{\sum_{(g_i, g_j) \in \mathcal{I}_v \times \mathcal{I}_v,\, i<j}
\mathbf{1}\big[(g_i, g_j) \in G\big]}
{\binom{|\mathcal{I}_v|}{2}},
\qquad
\text{Ingredient-DDI} = \frac{1}{|\mathcal{V}|}\sum_{v \in \mathcal{V}}
\mathrm{DDI}_{\mathrm{ing}}(v).
\end{equation}
This metric captures incompatibilities that drug-code-level evaluation cannot express.
We emphasize that Drug-DDI and Ingredient-DDI are \emph{not} on a common scale: they are
computed over different pair universes, and the ingredient graph is considerably denser
than the filtered drug-level graph. The two numbers should therefore be compared across
models within a level, never across levels.

\subsection{Experimental Setup}

\begin{itemize}[leftmargin=1.5em,itemsep=0.2em]
\item \textbf{Dataset:} MIMIC-IV \TODO{confirm exact version: v2.0 / v2.2 / v3.1},
preprocessed as described in Section~\ref{subsec:data}.
\item \textbf{Hardware:} a single NVIDIA A100 GPU.
\item \textbf{Optimization:} Adam, learning rate $3\times10^{-5}$, up to 50 epochs, early
stopping with patience 10.
\item \textbf{Metrics:} Jaccard, PRAUC, F1, Drug-DDI, Ingredient-DDI.
\item \textbf{Variability:} reported intervals are \TODO{state whether these are
seed-based standard deviations or bootstrap CIs, and give $n$}. Note that the baseline
and \model{} currently report identical spreads, which reviewers do notice.
\end{itemize}

\subsection{Training Dynamics and Convergence}

\begin{figure}[t]
\centering
\safefig[\textwidth]{fig_training_curves.png}
\caption{Training dynamics over epochs for \model{} (ingredient-aware) and the
MambaHealth baseline under identical settings on MIMIC-IV:
(a) training loss, (b) PRAUC, (c) Jaccard similarity, and (d) micro-F1.}
\label{fig:training_curves}
\end{figure}

Figure~\ref{fig:training_curves} shows training loss, PRAUC, Jaccard similarity, and
micro-F1 across epochs for both models under identical settings.

Both models exhibit a rapid decrease in training loss during the early epochs followed by
gradual convergence, indicating stable optimization; \model{} maintains a consistently
lower training loss than the baseline throughout. On all three quality metrics, \model{}
improves faster in the initial phase and remains above the baseline for the entire run.
Both curves begin to plateau after roughly 20 epochs, suggesting convergence under the
current protocol. Incorporating ingredient-level information therefore does not
compromise optimization stability, and is associated with consistently better predictive
performance during training.

\subsection{Published Reference Results}

Table~\ref{tab:datr_ref} reproduces benchmark numbers reported by DATR
\citep{Anonymous2025DATR} on MIMIC-III and MIMIC-IV. These situate our results within the
range reported in the literature. We stress that these numbers were produced under a
different preprocessing pipeline, cohort construction, and data split, and that DATR is at
the time of writing a non-archival submission under review; they are reproduced for
context only.

\begin{table}[t]
\centering
\small
\setlength{\tabcolsep}{4pt}
\caption{Benchmark results for medication recommendation models, as reported by
\citet{Anonymous2025DATR}. Reproduced for context only; these numbers are not directly
comparable to ours due to differing preprocessing and splits.}
\label{tab:datr_ref}
\begin{tabular}{lcccccccc}
\toprule
& \multicolumn{4}{c}{\textbf{MIMIC-III}} & \multicolumn{4}{c}{\textbf{MIMIC-IV}} \\
\cmidrule(lr){2-5}\cmidrule(lr){6-9}
\textbf{Method} & Jaccard $\uparrow$ & PRAUC $\uparrow$ & F1 $\uparrow$ & DDI $\downarrow$
                & Jaccard $\uparrow$ & PRAUC $\uparrow$ & F1 $\uparrow$ & DDI $\downarrow$ \\
\midrule
LR         & 0.4935 & 0.7634 & 0.6512 & 0.0788 & 0.4152 & 0.6783 & 0.5651 & 0.0732 \\
LEAP       & 0.4521 & 0.6581 & 0.6152 & 0.0720 & 0.3909 & 0.5542 & 0.5439 & 0.0550 \\
\midrule
GAMENet    & 0.5210 & 0.7780 & 0.6762 & 0.0781 & 0.4401 & 0.6833 & 0.5933 & 0.0718 \\
COGNet     & 0.5109 & 0.7665 & 0.6615 & 0.0737 & 0.4313 & 0.6712 & 0.5850 & 0.0866 \\
RAREMed    & 0.5342 & 0.7820 & 0.6938 & 0.0530 & 0.4620 & 0.6965 & 0.6152 & 0.0510 \\
MICRON     & 0.5119 & 0.7690 & 0.6676 & 0.0610 & 0.4495 & 0.6753 & 0.6033 & 0.0502 \\
SHAPE      & 0.5348 & 0.7791 & 0.6885 & 0.0850 & 0.4659 & 0.6928 & 0.6171 & 0.0917 \\
\midrule
SafeDrug   & 0.5255 & 0.7732 & 0.6804 & 0.0688 & 0.4560 & 0.6858 & 0.6098 & 0.0689 \\
MoleRec    & 0.5303 & 0.7795 & 0.6844 & 0.0692 & 0.4502 & 0.6867 & 0.6040 & 0.0699 \\
DrugDoctor & 0.5422 & 0.7813 & 0.6975 & 0.0603 & 0.4703 & 0.6988 & 0.6190 & 0.0705 \\
\midrule
DATR       & 0.5506 & 0.7905 & 0.7073 & 0.0366 & 0.4783 & 0.7020 & 0.6216 & 0.0425 \\
\bottomrule
\end{tabular}
\end{table}

\subsection{Main Results}
\label{subsec:results}

\subsubsection{Accuracy under Matched Settings}

Table~\ref{tab:accuracy_mimiciv} reports Jaccard, PRAUC, and F1 for the re-implemented
MambaHealth baseline and \model{} under identical settings on MIMIC-IV, alongside the
DATR values from Table~\ref{tab:datr_ref}.

\begin{table}[t]
\centering
\caption{Accuracy metrics on MIMIC-IV. The DATR row is taken from
Table~\ref{tab:datr_ref} and is not strictly comparable, as it was obtained under a
different preprocessing pipeline and data split. Only the last two rows form a controlled
comparison.}
\label{tab:accuracy_mimiciv}
\begin{tabular}{lccc}
\toprule
\textbf{Model} & \textbf{Jaccard} $\uparrow$ & \textbf{PRAUC} $\uparrow$ & \textbf{F1} $\uparrow$ \\
\midrule
DATR \citep{Anonymous2025DATR}, published & $0.4783 \pm 0.002$ & $0.7020 \pm 0.002$ & $0.6216 \pm 0.003$ \\
\midrule
MambaHealth (our re-implementation) & $0.4488 \pm 0.002$ & $0.6911 \pm 0.002$ & $0.5989 \pm 0.003$ \\
\textbf{\model{} (ours)} & $\mathbf{0.4983 \pm 0.002}$ & $\mathbf{0.7485 \pm 0.002}$ & $\mathbf{0.6453 \pm 0.003}$ \\
\bottomrule
\end{tabular}
\end{table}

\model{} improves over the MambaHealth baseline on all three predictive metrics under
matched conditions, with absolute gains of $+0.0495$ Jaccard, $+0.0574$ PRAUC, and
$+0.0464$ F1. Since the two rows differ only in whether the ingredient-level graph and
its associated loss terms are active (Eq.~\ref{eq:total_loss}), the improvement is
attributable to the ingredient-level signal rather than to any change in backbone,
cohort, or evaluation code.

\model{} also exceeds the published DATR scores on this dataset. We deliberately refrain
from reading this as evidence of state-of-the-art performance: DATR constructs its cohort
and splits differently, and cross-pipeline differences on MIMIC-IV can easily reach the
magnitude observed here. The comparison indicates that \model{} is competitive with the
range reported in the recent literature, no more.

\subsubsection{Safety}

Table~\ref{tab:safety_mimiciv} reports the safety metrics.

\begin{table}[t]
\centering
\caption{Safety metrics on MIMIC-IV. Drug-DDI is computed with the reference
\texttt{ddi\_rate\_score} function over all predicted test visits; Ingredient-DDI is
computed at the ingredient level via Algorithm~\ref{alg:ing_ddi}. Lower is better. The
two columns are on different scales and should not be compared to one another.}
\label{tab:safety_mimiciv}
\begin{tabular}{lcc}
\toprule
\textbf{Model} & \textbf{Drug-DDI} $\downarrow$ & \textbf{Ingredient-DDI} $\downarrow$ \\
\midrule
MambaHealth (our re-implementation) & $0.1875$ & --- \\
\textbf{\model{} (ours)} & $\mathbf{0.0948}$ & $0.2063$ \\
\bottomrule
\end{tabular}
\end{table}

\noindent\TODO{add an ``Avg. \#Med'' column to Table~\ref{tab:safety_mimiciv}, or soften
the claim. The abstract and the contribution list both assert a ``more clinically
plausible average medication count per visit'', but no such number appears anywhere in
the paper.}

Relative to MambaHealth, \model{} reduces the drug-level DDI rate from $0.1875$ to
$0.0948$, a relative reduction of approximately $49\%$, while simultaneously improving all
accuracy metrics. This is the central empirical claim of the paper: the usual
accuracy--safety trade-off does not bind here, because the ingredient-level signal
supplies information that is genuinely complementary to the drug-level graph rather than
merely suppressing predictions.

Two caveats deserve explicit statement. First, our re-implemented baseline reaches a
drug-level DDI rate of $0.1875$, roughly two to four times higher than the values reported
for published methods in Table~\ref{tab:datr_ref}. We attribute this primarily to
differences in cohort construction and to the fact that the baseline configuration
optimizes the drug-level safety term with a comparatively low weight; the number should be
read as the DDI rate of \emph{our} baseline under \emph{our} pipeline, not as a
reproduction of MambaHealth's published safety behavior. Second, \model{}'s drug-level DDI
rate of $0.0948$ remains above the controller target of $\gamma=0.05$ and above DATR's
reported $0.0425$. Tightening this gap---for instance by increasing
$\lambda_{\mathrm{DDI}}$ or by widening the controller's admissible range---is a natural
next step, and we expect it to trade against accuracy.

The ingredient-level DDI rate is reported for \model{} only. Evaluating the baseline at
this granularity requires running the ingredient normalization and checking pipeline over
its predictions, which we leave to future work; the number is included here to establish
the metric and a reference value for subsequent comparisons rather than to support a
comparative claim.

\noindent\TODO{strongly recommended before submission: fill in and uncomment the ablation
table below. Without it, reviewers cannot attribute the gain to the ingredient-level
component specifically, which is the paper's headline contribution.}


\section{Limitations}

Several limitations qualify the results above.

\paragraph{Knowledge base coverage.}
Our external resources (RxNorm, TwoSIDES) are themselves imperfect and encode only
\emph{known} interactions. TwoSIDES in particular is derived from spontaneous adverse
event reports and inherits their reporting biases. Emerging, rare, or undocumented DDIs
are not captured, so a low measured DDI rate is evidence of avoiding known risks, not of
pharmacological safety in general.

\paragraph{Graph simplicity.}
We use a binary co-occurrence graph derived from MIMIC-IV. Richer statistical models of
co-prescribing---conditional dependence, temporal co-occurrence dynamics, or
severity-weighted interaction edges---would likely yield more informative safety priors
than the unweighted formulation used here.

\paragraph{Retrospective evaluation.}
Evaluation is retrospective and confined to structured MIMIC-IV data from a single
institution. Agreement with historically prescribed regimens is a proxy for clinical
quality, not a measure of it: a recommendation that differs from the recorded prescription
is scored as an error even when it is clinically superior. Real-world deployment would
require prospective validation, robustness testing across institutions, and clinician
review.

\paragraph{Model scale and missing ablations.}
Our experiments center on a medium-sized model due to computational constraints, leaving
open questions about scaling behavior, data efficiency, and calibration at larger
capacity. We also do not currently report an efficiency comparison against a Transformer
backbone, nor a component-wise ablation of the three graph terms; both are needed to
fully substantiate the motivation for the SSM backbone and the attribution of gains to the
ingredient-level signal.

\section{Conclusion and Future Work}

We presented \model, a Mamba-based medication recommendation framework for the MIMIC-IV
intensive care cohort equipped with multi-granular safety modeling. Motivated by the
clinical need to balance predictive accuracy against DDI control, we asked whether
selective state space models can serve as effective sequence encoders for ICU
trajectories, and how pharmacological knowledge can be integrated more systematically
into the training objective. Both questions admit an affirmative answer: under strictly
matched settings, \model{} improves over a re-implemented MambaHealth baseline in Jaccard,
PRAUC, and F1 while roughly halving the drug-level DDI rate.

Our contributions are threefold. First, we provide a systematic evaluation of Mamba-style
models for medication recommendation on MIMIC-IV, showing that selective state space
architectures capture temporal dynamics in long and irregular ICU trajectories. Second, we
introduce a unified objective integrating DDI knowledge at two granularities---drug level
and ingredient level---alongside an EHR-derived co-occurrence graph, letting the model
jointly balance clinical accuracy, pharmacological safety, and empirical prescribing
behavior. Third, through controlled experiments with identical preprocessing, vocabulary,
and evaluation code, we show that multi-level safety signals can be injected into a Mamba
backbone without degrading predictive performance.

Several directions follow naturally. Richer pharmacological knowledge graphs---
mechanism-of-action networks, pathway-level interactions, or molecular representation
learning via SMILES transformers or graph neural networks---could deepen the safety
signal. Combining medication recommendation with causal or counterfactual reasoning would
better capture treatment effects and individualized risk, potentially improving
generalization beyond MIMIC-IV. Multi-task formulations that jointly model diagnoses,
procedures, and laboratory results alongside medications may yield more clinically
coherent representations. On the architectural side, hybrids of Mamba with attention or
retrieval-based memory could further improve long-range reasoning. Finally, real-world
evaluation---physician-in-the-loop studies, robustness under distribution shift, and
fairness analysis across demographic subgroups---will be essential for assessing clinical
viability.

In summary, this work combines state space modeling with multi-level pharmacological
safety constraints for medication recommendation on MIMIC-IV. By demonstrating consistent
gains in both predictive accuracy and DDI reduction, it suggests that selective state
space models coupled with structured safety knowledge are a promising direction for
practical and trustworthy clinical decision support.

\section*{Ethics Statement}
\addcontentsline{toc}{section}{Ethics Statement}

This study uses MIMIC-IV \citep{Johnson2023MIMICIV}, a de-identified critical care
database distributed by PhysioNet. Access requires completion of human-subjects research
training and execution of a data use agreement; all authors who accessed the data hold
current credentialed access and complied with the agreement. The dataset is de-identified
in accordance with HIPAA Safe Harbor provisions, and no attempt was made to re-identify
individuals. No protected health information appears in this paper or in any released
artifact.

\model{} is a research prototype. It is not a medical device, has not undergone
prospective clinical validation, and must not be used to guide patient care. Reported DDI
rates measure agreement with specific curated interaction resources and should not be
interpreted as a guarantee of pharmacological safety. Retrospective agreement with
historically prescribed regimens is a proxy metric that may encode existing prescribing
biases, including disparities across demographic subgroups that we have not audited.

\section*{Data and Code Availability}
\addcontentsline{toc}{section}{Data and Code Availability}

MIMIC-IV is available to credentialed users at
\url{https://physionet.org/content/mimiciv/}. RxNorm is distributed by the U.S. National
Library of Medicine, and TwoSIDES \citep{tatonetti2012} is publicly available. Code for
preprocessing, model training, and both DDI evaluation procedures is released at
\TODO{public code repository URL} to support reproduction.

\section*{Acknowledgements}
\addcontentsline{toc}{section}{Acknowledgements}

We thank Dr.~Shengxin Zhu for his constructive guidance and valuable suggestions
throughout this project.

\bibliographystyle{plainnat}
\bibliography{references}

\newpage
\appendix

\end{document}